\documentclass[longbibliography,noeprint,noshowpacs,nopreprintnumbers,twocolumn,pra,showpacs,superscriptaddress,amsmath,amssymb,citeautoscript,aps,10pt]{revtex4-2}

\usepackage{graphicx}
\usepackage{dcolumn}
\usepackage{color}
\usepackage{bm}
\usepackage[hidelinks]{hyperref}
\hypersetup{
    colorlinks,
    citecolor=blue,
    filecolor=blue,
    linkcolor=blue,
    urlcolor=blue
}
\graphicspath{{Figures/}{./Figures/}{.}}

\DeclareMathOperator{\tr}{tr}
\DeclareMathOperator\arctanh{arctanh}
\DeclareMathOperator\re{Re}
\DeclareMathOperator\im{Im}

\begin{document}

\title{The dynamical bulk boundary correspondence and dynamical quantum phase transitions in the Benalcazar-Bernevig-Hughes model}

\author{Tomasz Mas{\l}owski}
\affiliation{The Faculty of Mathematics and Applied Physics, Rzesz\'ow University of Technology, al.~Powsta\'nc\'ow Warszawy 6, 35-959 Rzesz\'ow, Poland}
\author{Nicholas Sedlmayr}
\email[e-mail:]{sedlmayr@umcs.pl}
\affiliation{Institute of Physics, M. Curie-Sk\l{}odowska University, 20-031 Lublin, Poland}

\date{\today}

\begin{abstract}
In this article we demonstrate that dynamical quantum phase transitions occur for an exemplary higher order topological insulator, the Benalcazar-Bernevig-Hughes model, following quenches across a topological phase boundary. A dynamical bulk boundary correspondence is also seen both in the eigenvalues of the Loschmidt overlap matrix and the boundary return rate. The latter is found from a finite size scaling analysis for which the relative simplicity of the model is crucial. Contrary to the usual two dimensional case the dynamical quantum phase transitions in this model show up as cusps in the return rate, as for a one dimensional model, rather than as cusps in its derivative as would be typical for a two dimensional model. We explain the origin of this behaviour.
\end{abstract}

\maketitle

\section{Introduction}\label{sec:intro}

In the last ten years an analogue of quantum phase transitions which can occur in time following sudden quenches has been developed. These dynamical quantum phase transitions (DQPTs)~\cite{Heyl2013,Heyl2018a,Sedlmayr2019a} have become one method for systematically studying the non-equilibrium behaviour of a wide variety of quantum systems. Although initial studies suggested a close connection between the equilibrium phase diagram and DQPTs, interestingly in general no such connection holds~\cite{Vajna2014,Andraschko2014,Vajna2015,Karrasch2017,Jafari2017,Jafari2017a,Cheraghi2018,Jafari2019,Wrzesniewski2022} allowing DQPTs to be a window into genuinely non-equilibrium phenomena. Following the introduction of the concept a large amount of theoretical work has followed~\cite{Heyl2013,Karrasch2013,Sharma2014,Heyl2014,Heyl2015,Sharma2015,Halimeh2017,Homrighausen2017,Halimeh2018,Shpielberg2018,Sedlmayr2018,Sedlmayr2018b,Zunkovic2018,Yang2019,Srivastav2019,Huang2019,Gurarie2019,Abdi2019,Maslowski2020,Puebla2020,Link2020,Sun2020,Rylands2020,Trapin2021,Yu2021,Halimeh2021,Halimeh2021a,DeNicola2021,Cheraghi2021,Cao2021,Bandyopadhyay2021,Maslowski2023,Cheraghi2023}, along with several experiments on ion trap, cold atom, and quantum simulator platforms~\cite{Jurcevic2017,Flaschner2018,Zhang2017b,Guo2019,Smale2019,Nie2020,Tian2020}. Amongst other developments extensions to finite temperatures and open or dissipative systems have been made~\cite{Mera2017,Sedlmayr2018b,Bhattacharya2017a,Heyl2017,Abeling2016,Lang2018,Lang2018a,Kyaw2020,Starchl2022,Naji2022,Kawabata2022}. Many studies remain focused on spin chains and one dimension, though multi-band models~\cite{Huang2016,Jafari2019,Mendl2019,Maslowski2020,Maslowski2023}, and to higher dimensional systems~\cite{Vajna2015,DeNicola2022,Hashizume2022,Brange2022,Maslowski2023,Kosior2023} have also been considered. Connections have also been considered between DQPTs and other phenomena, for example the entanglement entropy~\cite{Sedlmayr2018}, string order parameters~\cite{Uhrich2020}, the characteristic function of work~\cite{Abeling2016,Talkner2007}, crossovers in the quasiparticle spectra~\cite{Halimeh2018}, and out of time ordered correlators~\cite{Heyl2018,Chen2020,Zamani2022,Sedlmayr2023}.

DQPTs have been shown to occur in many different examples of topological matter~\cite{Vajna2015,Schmitt2015,Jafari2016,Jafari2017a,Sedlmayr2018,Jafari2018,Zache2019,Maslowski2020,Okugawa2021,Maslowski2023} which is also a recent growth industry~\cite{Hasan2010,Chiu2016,Wen2017}. One of the interesting phenomena seen in topological materials is the relation between the bulk topology and protected edge states of one dimension lower~\cite{Teo2010}, this is referred to as the bulk-boundary correspondence. In a higher order topological insulator the edge modes have a dimension lower than the bulk by more than one~\cite{Volovik2010,Sitte2012,Zhang2013b,Benalcazar2017,Benalcazar2017a,Langbehn2017,Song2017,Schindler2018,Fang2019,Trifunovic2018,Trifunovic2021,Xie2021}. Dynamical order parameters for DQPTs have been found~\cite{Budich2016,Heyl2017,Bhattacharya2017a,Wang2019e} and a dynamical bulk boundary correspondence has also been seen~\cite{Sedlmayr2018,Maslowski2020}, including in higher order topological matter~\cite{Maslowski2023}.

In this work we focus on a paradigmatic example of a two dimensional higher order topological insulator: the Benalcazar-Bernevig-Hughes (BBH) model~\cite{Benalcazar2017,Benalcazar2017a}. This allows us to derive expressions which determine the DQPTs analytically, and obtain numerical solutions sufficient for performing a finite sized scaling analysis. We show that this two dimensional model can also exhibit behaviour characteristic of one dimensional DQPTs.

In section \ref{sec:dqpt} we introduce the concept of DQPTs and the methods for calculations, and in section \ref{sec:mod} we introduce the BBH model, describing its symmetry properties, spectra, and topological phase diagram. Section \ref{sec:dqpt} contains the results on DQPTs and the dynamical bulk-boundary correspondence, following which we conclude.

\section{Dynamical Quantum Phase Transitions}\label{sec:dqpt}

DQPTs are defined using the overlap between an initial state $|\Psi_0\rangle$ and this state time evolved by a Hamiltonian $\mathcal{H}^1$. This overlap is called the Loschmidt echo~\cite{Heyl2013}
\begin{equation}
L(t)=\langle\Psi_0|e^{-i\mathcal{H}^1t}|\Psi_0\rangle\,.
\end{equation}
For complex $t$ the boundary part of $L(t)$ at $\mathrm{Im}(t)\to\infty$ is equivalent to the standard partition function. This corresponds to a quench scenario where one can consider the initial state as the ground state of a Hamiltonian $\mathcal{H}^0$ which is then suddenly changed and the system is time evolved with a different Hamiltonian $\mathcal{H}^1$. Properties of the time evolution can therefore be related to the properties of $\mathcal{H}^0$ and $\mathcal{H}^1$.

One can then define a ``free energy'' called the return rate
\begin{equation}
   l_0(t)=\lim_{N\to\infty} l_N(t)=-\lim_{N\to\infty}\frac{1}{N}\ln\left|L(t)\right|\,,
\end{equation}
which has non-analyticities at zeroes of the Loschmidt echo. In analogy to a standard quantum phase transition these non-analyticities are referred to as dynamical quantum phase transitions. In one dimensional systems the non-analyticities occur at critical times when the zeroes of $L(t)$ in the complex plane cross the real time axis~\cite{Heyl2013}, known as Fisher zeroes. In the bulk the line of Fisher zeroes can be parameterised by momenta. At a critical momenta the line can cross the real axis and a DQPT occurs. In two dimensions the situation is more complicated as the Fisher zeroes now form a plane and the critical region which crosses the real axis is extended over a finite range of time~\cite{Vajna2015}. W=In that case where the density of these zeroes diverges a cusp forms in the derivative of the return rate.

For our purposes a convenient representation for the Loschmidt echo is~\cite{Levitov1996,Klich2003,Rossini2007,Sedlmayr2018}
\begin{equation}\label{rle}
L(t)=\det\,\underbrace{\left[1-\mathbf{\mathcal{C}}+\mathbf{\mathcal{C}}e^{i {\bm H}^1 t}\right]}_{\equiv {\bm M}(t)}
\end{equation}
where ${\bm H}^1$ is the Hamiltonian matrix and $\mathbf{\mathcal{C}}$ is the correlation matrix $\mathcal{C}_{ij}=\langle\Psi_0|c_i^\dagger c_j|\Psi_0\rangle$ for some complete basis set of creation operators $\{c^\dagger_j\}$. In terms of the eigenvalues $\lambda_i(t)$ of ${\bm M}(t)$ one finds
\begin{equation}\label{ele}
L(t)=\prod_i\lambda_i(t)\,,
\end{equation}
and
\begin{equation}\label{rre}
l_N(t)=-\frac{1}{N}\sum_i\ln\left|\lambda_i(t)\right|\,.
\end{equation}
We also define the required derivative
\begin{equation}\label{retderdef}
d_N(t)\equiv\dot{l}_N(t)=-\frac{1}{N}\sum_{i}\left|\frac{\dot{\lambda}_i(t)}{\lambda_i(t)}\right|\,,
\end{equation}
which in terms of the Loschmidt matrix is
\begin{equation}\label{retder}
d_N(t)\equiv\dot{l}_N(t)=-\frac{1}{N}\,\mathrm{Re}\left(\tr\left[\dot{\bm M}(t){\bm M}^{-1}(t)\right]\right)\,.
\end{equation}
In the thermodynamic limit we write $d_0(t)=\lim_{N\to\infty}d_N(t)$.

In equilibrium the bulk-boundary correspondence relates the bulk topology to the existence of edge modes~\cite{Teo2010}. For DQPTs a dynamical bulk boundary correspondence has been discovered which relates the change in topology between the initial state and the time evolving Hamiltonian to boundary contributions to the return rate~\cite{Sedlmayr2018,Maslowski2020,Maslowski2023}. The boundary return rate in one dimension, $l^{1D}_{B}(t)$, can be found from
\begin{equation}\label{bbreturn1d}
l_N^{1D}(t)\sim l^{1D}_0(t)+\frac{l^{1D}_{B}(t)}{N}\,.
\end{equation}
In the simplest scenario a quench from the topologically non-trivial to the topologically trivial case results in periodically appearing and vanishing plateaus in $l^{1D}_{B}(t)$ between critical times. These plateaus can be directly related to zero eigenvalues of the Loschmidt matrix which become pinned to zero between alternating critical times, when the spectrum of the Loschmidt matrix becomes gapless, thus demonstrating the close analogy to the equilibrium bulk-boundary correspondence. Indeed the number of zero eigenvalues is related to the topological indices of the initial and time evolving Hamiltonians, though this connection is not necessarily that direct when larger topological indices are involved~\cite{Sedlmayr2018} or for cases where DQPTs can occur for quenches within a topologically non-trivial phase~\cite{Maslowski2020}.

In two dimensions a similar dynamical bulk boundary correspondence has been seen in intrinsic and extrinsic higher order topological insulators~\cite{Maslowski2023}. In that case the critical times become extended into regions of finite duration, but pinned zero modes of the Loschmidt matrix can still be seen between successive critical regions. In principle one would expect also plateaus in an appropriately defined boundary return rate would also occur. However the models previously studied were too complex for good enough data to be produced to determine this. One principle goal of this work is to fill this gap by focusing on a minimal model of a higher order topological insulator. If $N$ is the total number of atoms in the two dimensional lattice then we may expect scaling of the form
\begin{equation}\label{bbreturn}
l_N(t)\sim l_0(t)+\frac{l_{B}(t)}{\sqrt{N}}\,.
\end{equation}
The boundary contribution at a definite system size can be directly compared to the contribution from the $\tilde n$ eigenvalues $\lambda_i(t)$ which become pinned to zero:
\begin{equation}\label{lapprox}
    l_N(t)-l_0(t)\approx -\frac{1}{N}\sum_{n=0}^{\tilde n-1}\ln\left|\lambda_n(t)\right|\,.
\end{equation}
For the model we will focus on here the topological regime contains four corner states and we find that $\tilde n=4$.

\section{The Benalcazar-Bernevig-Hughes Model}\label{sec:mod}

\begin{figure}
    \centering
    \includegraphics[width=0.95\columnwidth]{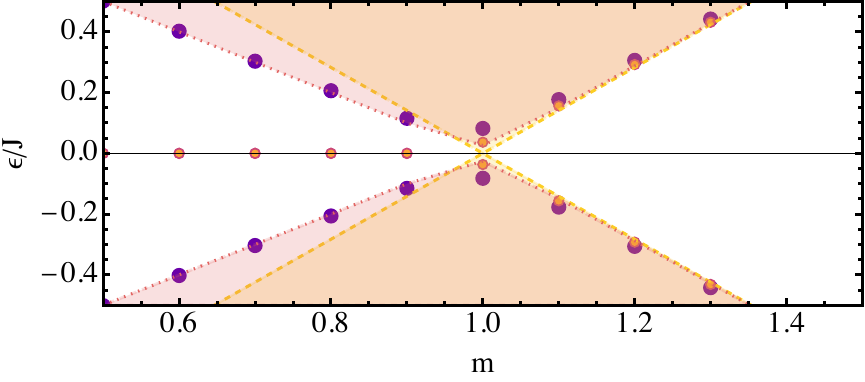}
    \caption{The spectrum of the BBH model as a function of $m$. Shown are the lowest 6 eigenenergies for the square lattice with open boundary conditions (circles) for $N=60^2$. The bulk gap is shown as a shaded orange region, and the gapped one dimensional edge modes on the edge of the square lattice are shown as the purple shaded region.}
    \label{fig:spectra}
\end{figure}

The Benalcazar-Bernevig-Hughes (BBH) model is a minimal four band model given by 
\begin{equation}\label{ham}
\mathcal{H}_{m}
= J\vec{\Gamma} \cdot\vec{d}^m_{\vec{k}}\,,
\end{equation}
where $\vec{\Gamma}$ is a vector containing four $4\times4$ matrices. The matrices are given by $\Gamma_k=-\tau_2\sigma_k$ for $k=1,2,3$, and by $\Gamma_4=\tau_1\sigma_0$. The momentum dependent vector defining the Hamiltonian is
\begin{equation}  
  \vec{d}^m_{\vec{k}} =
    \begin{pmatrix}
       \sin k_y\\
       m+\cos k_y\\
       \sin k_x\\
       m+\cos k_x
    \end{pmatrix}\,.
\end{equation}
$J$ is an overall energy scale of the hopping terms and we will set everywhere $J=1$ and $\hbar=1$. This is an intrinsic higher order topological insulator with four corner modes, see figure \ref{fig:spectra} for the spectrum as a function of $m$. The bulk eigenenergies are two-fold degenerate and are given by
\begin{equation}
    \pm\epsilon^{m}_{kx,ky}=\pm \sqrt{2}\sqrt{1+m^2+m\left(\cos k_x+\cos k_y\right)}\,.
\end{equation}

This model has several symmetries. First a global particle hole symmetry $\mathcal{C}=\tau_3\times\sigma_0\hat K$ satisfying $\{\mathcal{C},\mathcal{H}_{m}\}=0$ and $\mathcal{C}^2=1$. There is also a ``time-reversal'' symmetry $\mathcal{T}=\tau_0\times\sigma_0\hat K$ which satisfies $\{\mathcal{T},\mathcal{H}_{m}\}=0$ and $\mathcal{T}^2=1$, with $\hat K$ being charge conjugation. Finally there are also crystalline symmetries present, such as the mirror symmetries~\cite{Benalcazar2017a,Trifunovic2021}
\begin{equation}
    \mathcal{U}_y\mathcal{H}_{m}(-k_x,k_y)\mathcal{U}_y^\dagger=\mathcal{H}_{m}(k_x,k_y)
\end{equation}
and
\begin{equation}
    \mathcal{U}_x\mathcal{H}_{m}(k_x,-k_y)\mathcal{U}_x^\dagger=\mathcal{H}_{m}(k_x,k_y)\,.
\end{equation}
Here $\mathcal{U}_y=\tau_1\sigma_3$ and $\mathcal{U}_x=\tau_1\sigma_1$. The other crystalline symmetry is a four fold rotational symmetry
\begin{equation}
    \mathcal{U}_4\mathcal{H}_{m}(-k_y,k_x)\mathcal{U}_4^\dagger=\mathcal{H}_{m}(k_x,k_y)
\end{equation}
with
\begin{equation}
    \mathcal{U}_4=\begin{pmatrix}
    0&0&1&0\\
    0&0&0&1\\
    0&-1&0&0\\
    1&0&0&0
    \end{pmatrix}
\end{equation}
and $\mathcal{U}_4^4=-1$. 
The crystalline symmetries become broken at the edges of the model, gapping the one dimensional edge modes and resulting in the corner modes. Due to the four fold rotational symmetry it is clear that there must be four corner modes for this model.

For all examples shown throughout this article we choose $m=0.5$ for the topologically non-trivial phase and $m=1.5$ for the topologically trivial phase. No results depend qualitatively on the exact values used. For ease of reference we label the topologically non-trivial phase by an invariant $\nu=1$ and the topologically trivial phase by an invariant $\nu=0$.

\section{Results}

From equation \eqref{rle} one can readily derive the bulk expression for the Loschmidt matrix for the system with periodic boundary conditions. For a quench from $\mathcal{H}^0=\mathcal{H}_m$ to $\mathcal{H}^1=\mathcal{H}_{m'}$
\begin{equation}\label{loexp}
    L(t)=\prod_{\vec{k}}\left[\cos\left(\epsilon^{m'}_{\vec{k}}t\right)+i\cos\delta\phi_{\vec{k}}\sin\left(\epsilon^{m'}_{\vec{k}}t\right)\right]^2\,,
\end{equation}
where
\begin{equation}
    \cos\delta\phi_{\vec{k}}=-\frac{\vec{d}^m_{\vec{k}}\cdot\vec{d}^{m'}_{\vec{k}}}{\epsilon^{m}_{\vec{k}}\epsilon^{m'}_{\vec{k}}}\,.
\end{equation}
This is closely related to the standard expression for a two band topological insulator~\cite{Vajna2015}, but we note is not a general expression for a four band model~\cite{Maslowski2020,Maslowski2023}.

The first condition for the critical times to occur is for $\cos\delta\phi_{\vec{k}}=0$ which happens for the critical momenta satisfying
\begin{equation}\label{critm}
    \cos k^*_x+\cos k^*_y=-2\frac{1+mm'}{m+m'}\,.
\end{equation}
This can be solved for real momenta only if
\begin{equation}\label{criteria}
    m(1-m')>(1-m')\,.
\end{equation}
Hence for $m'>1$ one needs $m<1$ and vice versa. These are precisely those quenches which cross the equilibrium phase boundary, as one would expect for a simple two band topological insulator~\cite{Vajna2015}. The critical times are then, for $n=0,1,2,\ldots$, given by
\begin{equation}\label{critt}
    t_c=\frac{\pi(2n+1)}{2\epsilon^{m'}_{k^*_x,k^*_y}}=\frac{\pi(2n+1)}{2\sqrt{2}} \sqrt{\frac{m+m'}{\left({m'}^2-1\right) (m'-m)}}\,.
\end{equation}
In this case, because the condition for the momenta \eqref{critm} appears as it does in the energy, the plane of Fisher zeroes collapses to a line, and there is a single critical time as in one dimension. From \eqref{critt} it is clear that the Fisher zeroes will only cross the real axis when either $m'>1$ and $m'>m$, or when $m'<1$ and $m'<m$, assuming that both $m'$ and $m$ are positive.

\begin{figure}
    \centering
    \includegraphics[width=0.45\columnwidth]{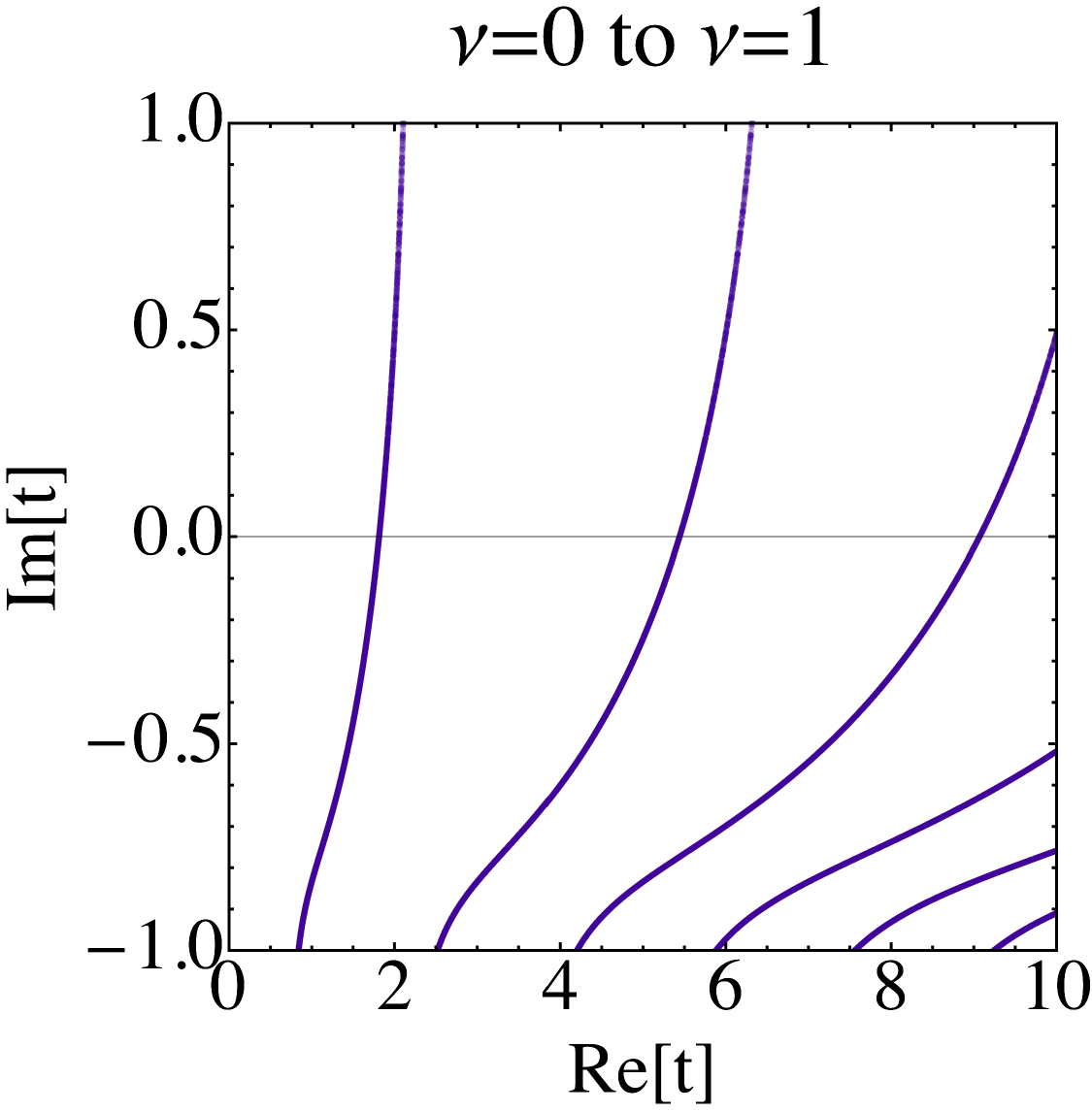}
    \includegraphics[width=0.45\columnwidth]{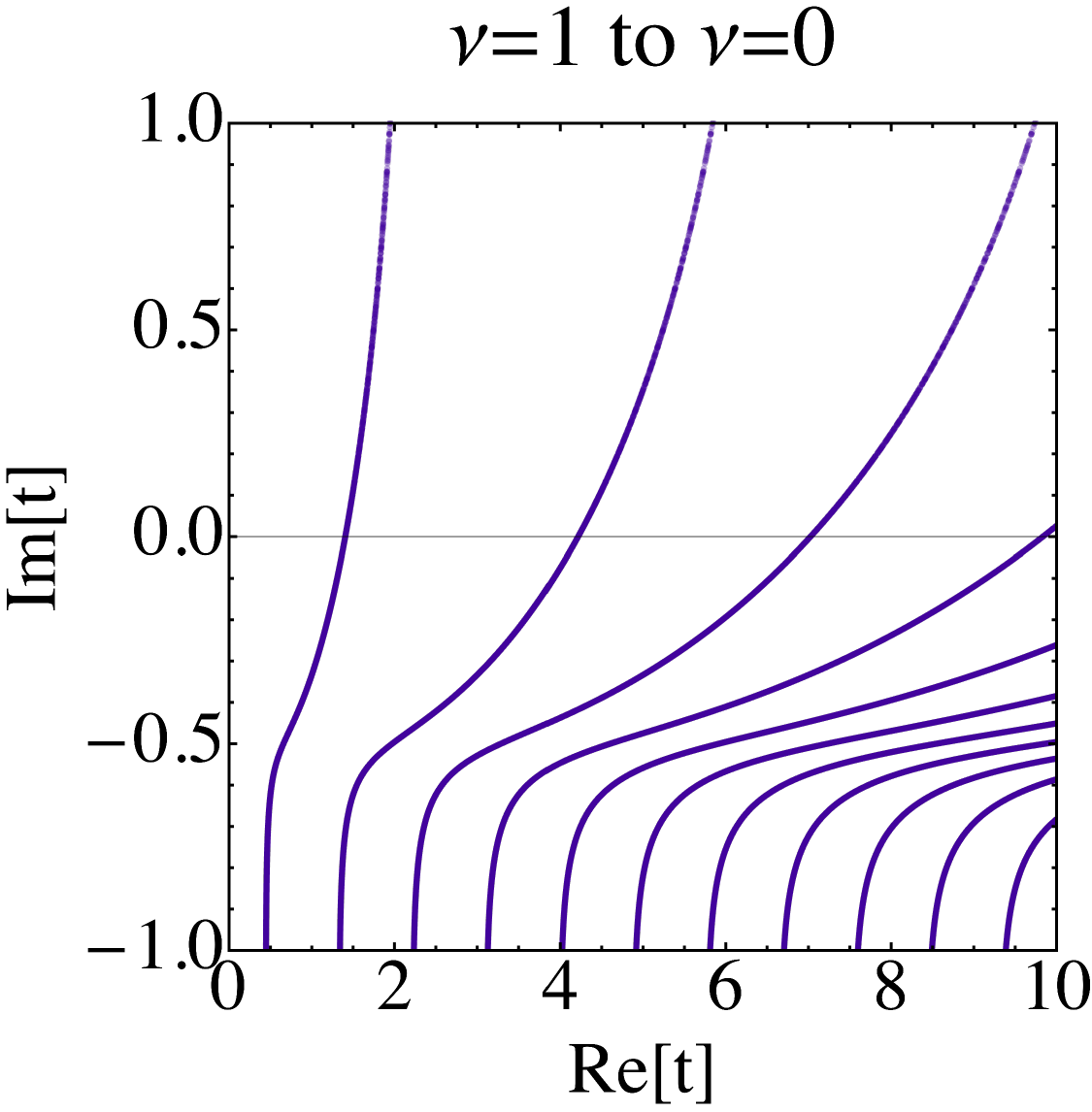}
    \caption{The Fisher zeroes in the complex $t$ plane for quenches across the equilibrium topological phase boundary. Critical times when the zeroes cross the real axis are clearly visible. In this case each line is parameterised by both $k_x$ and $k_y$ and so there are many zeroes at each point. The quench $\nu:0\to1$ is for $m=1.5$ and $m'=0.5$. The quench $\nu:1\to0$ is for $m=0.5$ and $m'=1.5$.}
    \label{fig:fisher}
\end{figure}

The Fisher zeroes themselves can also be easily found from \eqref{loexp}:
\begin{eqnarray}
\re[t]&=&\frac{\pi(2n+1)}{2\epsilon^{m'}_{\vec{k}}}\,, \\
\im[t]&=&\frac{\arctanh\left[\cos\delta\phi_{\vec{k}}\right]}{\epsilon^{m'}_{\vec{k}}} \,,
\end{eqnarray}
see figures \ref{fig:fisher} for examples. If the condition \eqref{criteria} is not met then the Fisher zeroes do not cross the real time axis. Here we show exemplary quenches which cross the equilibrium phase boundary in the two different directions.

\begin{figure}
    \centering
    \includegraphics[width=0.95\columnwidth]{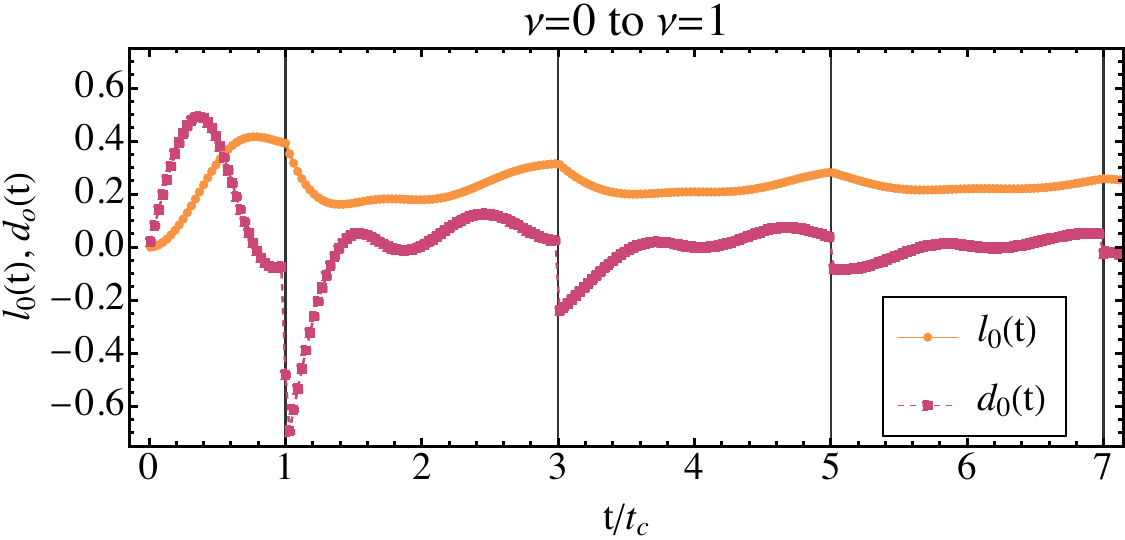}\\
    \includegraphics[width=0.9\columnwidth]{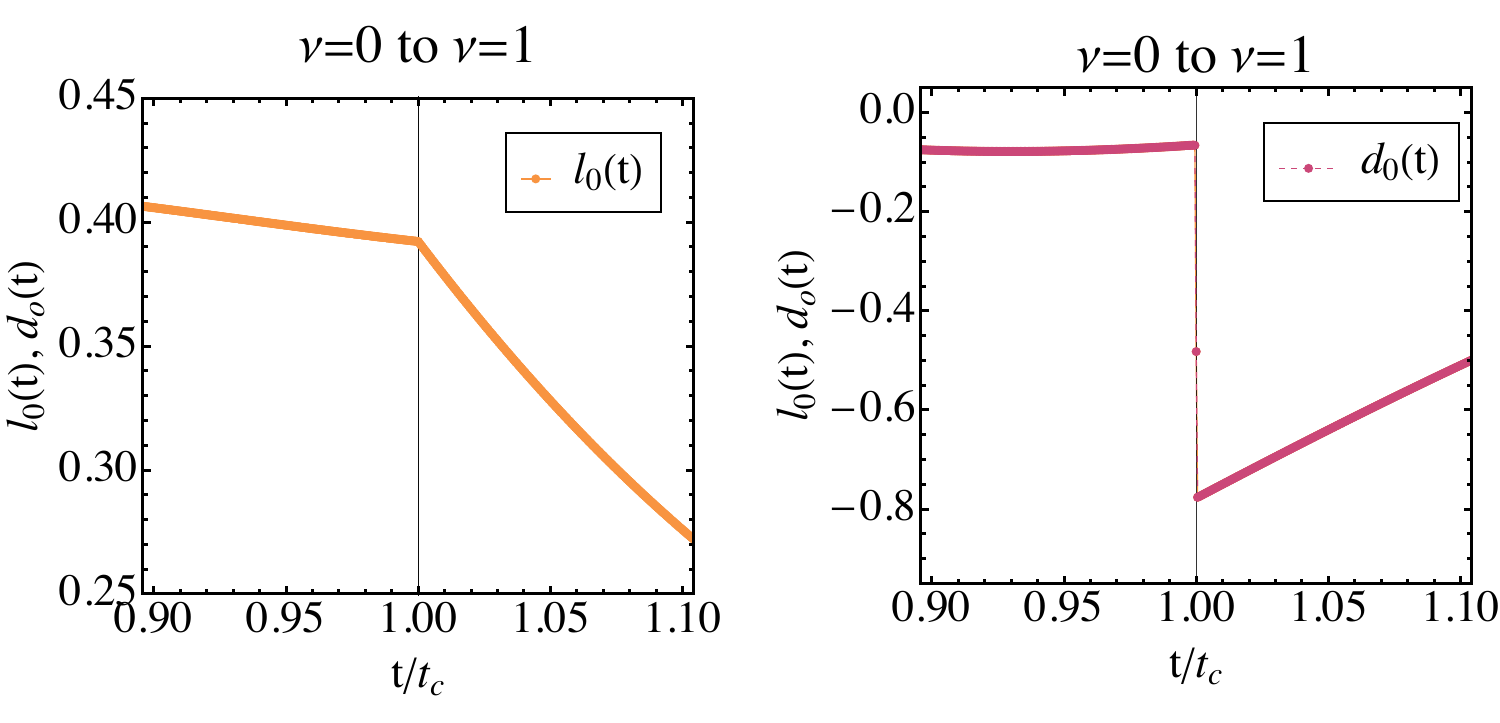}
    \caption{The return rate and its derivative for a quench from the topologically trivial to the topologically non-trivial phase. A jump in $d_0(t)$ at $t_c$ is clearly visible, as well as a cusp in the return rate itself. The lower panels show a zoom in for the first critical time, demonstrating the cusp in the return rate and the discontinuity in its derivative. The quench parameters are $m=1.5$ and $m'=0.5$.}
    \label{fig:rr01}
\end{figure}

\begin{figure}
    \centering
    \includegraphics[width=0.95\columnwidth]{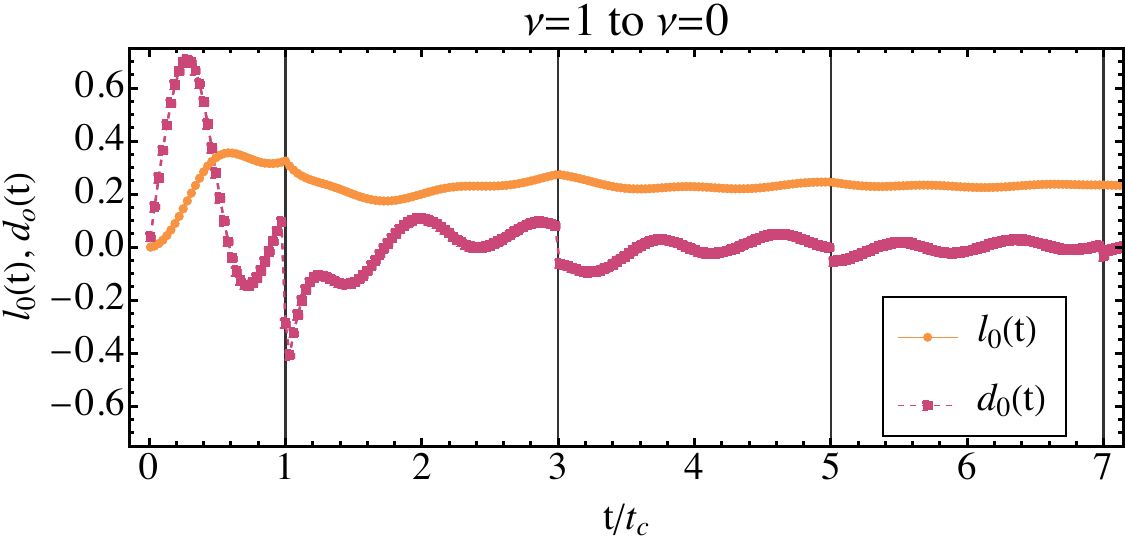}\\
    \includegraphics[width=0.9\columnwidth]{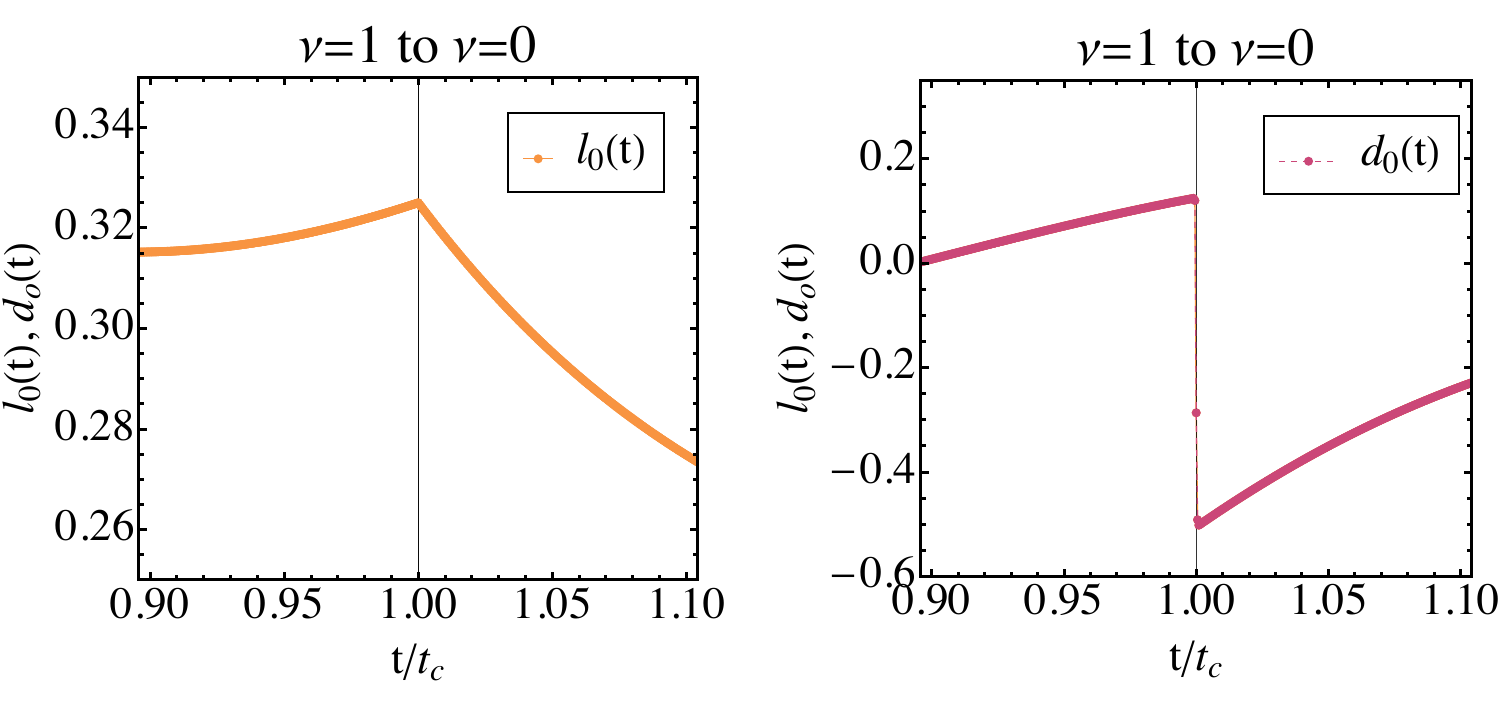}
    \caption{The return rate and its derivative for a quench from the topologically non-trivial to the topologically trivial phase. A jump in $d_0(t)$ at $t_c$ is clearly visible, as well as a cusp in the return rate itself. The lower panels show a zoom in for the first critical time, demonstrating the cusp in the return rate and the discontinuity in its derivative. The quench parameters are $m=0.5$ and $m'=1.5$.}
    \label{fig:rr10}
\end{figure}

For the derivative of the return rate we expect a sudden jump at the critical times, which can be seen in figures \ref{fig:rr01} and \ref{fig:rr10} for the quenches from topologically trivial to non-trivial and vice versa. This corresponds to a cusp in the return rate itself as is seen in one dimensional models. As the entire area of Fisher zeroes is collapsed onto a single curve for the BBH model this is to be expected.  For the bulk there is no qualitative difference between these two possible quench scenarios, though specific details do of course change.

\subsection{The Dynamical Bulk-Boundary Correspondence}

\begin{figure}
    \centering
    \includegraphics[width=0.95\columnwidth]{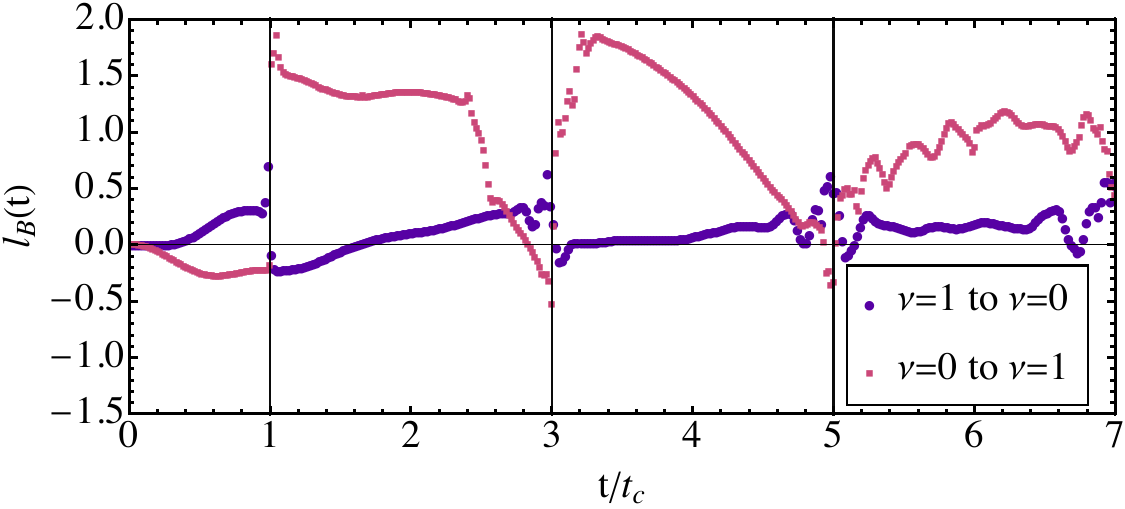}
    \caption{The boundary return rate $l_B(t)$ extracted from a finite size scaling analysis for system sizes $\sqrt{N}\in\{30,35,40,45,50,55,60\}$, see equation \eqref{bbreturn}. As expected from the dynamical bulk-boundary correspondence the quench into the non-trivial phase shows large features between successive critical times. For a detailed discussion see the main text.}
    \label{fig:scaling}
\end{figure}

We now turn to the boundary contributions. In figure \ref{fig:scaling} we show the boundary return rate $l_B$ extracted from a finite scaling analysis for system sizes $\sqrt{N}\in\{30,35,40,45,50,55,60\}$, see equation \eqref{bbreturn}. $N=60^2$ is the largest system size we were able to reach, placing some limitations on the scaling analysis. We recall that in a one-dimensional topological system the dynamical bulk-boundary effect corresponds to a plateau forming for quenches into the topologically non-trivial phase~\cite{Sedlmayr2018,Maslowski2020}. This plateau appears and disappears between successive critical times. For the opposite quench direction only small fluctuations in $l_B(t)$ occur. Similarly here we find only relatively small fluctuations in $l_B(t)$ for the quench from $\nu:1\to0$, see figure \ref{fig:scaling}. For the quench from $\nu:0\to1$ a larger plateau-like structure can be seen to form between the first two critical times. However between the next critical times it is slow to decay, and already after the third critical time it is less clear, though $l_B(t)$ remains larger for the quench to the topologically non-trivial phase compared to the quench into the trivial phase as predicted. The origin of the plateau can be traced to the Loschmidt matrix eigenvalues, it is caused by zero eigenvalues which appear and disappear between successive critical times when the gap in the Loschmidt matrix spectrum closes~\cite{Sedlmayr2018,Maslowski2020}. It is this behaviour which is referred to as the dynamical bulk-boundary correspondence. To clarify the situation here we now turn to the eigenvalues of the Loschmidt matrix. 

\begin{figure}
    \centering
    \includegraphics[width=0.95\columnwidth]{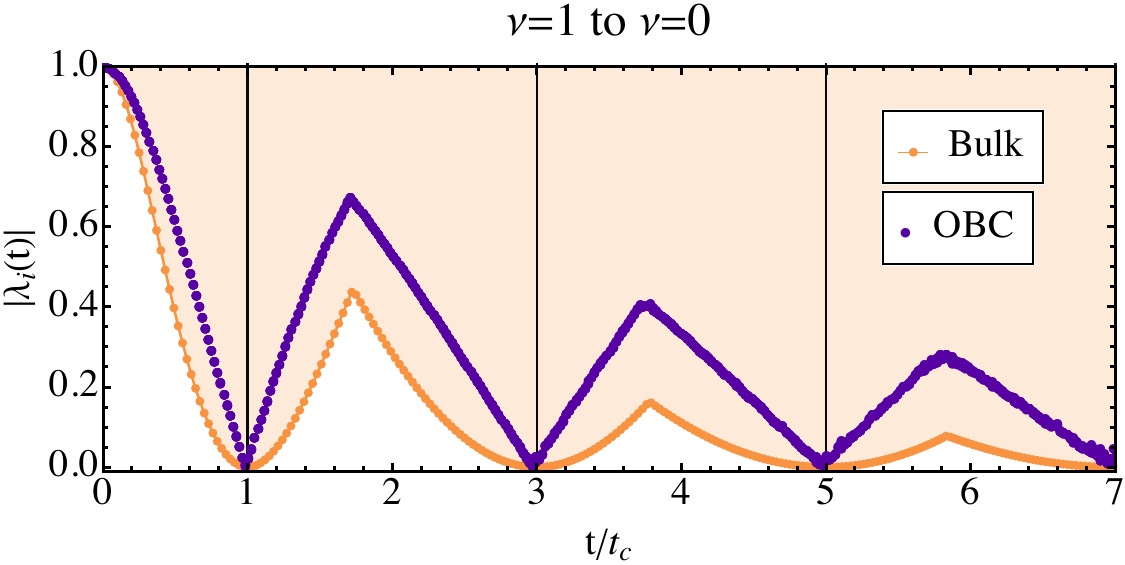}\\
    \includegraphics[width=0.95\columnwidth]{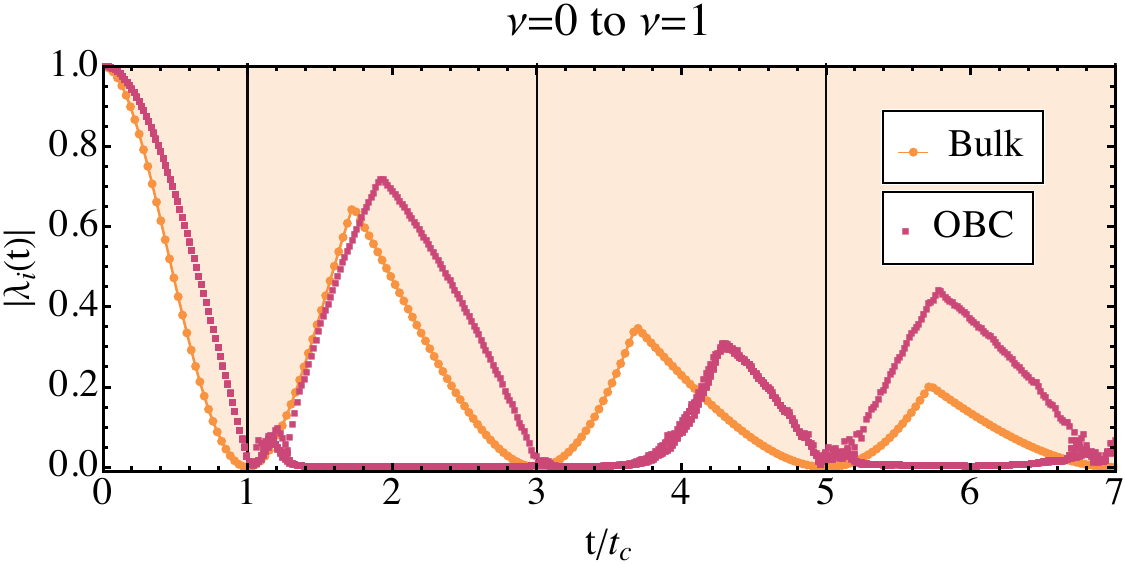}
    \caption{The eigenvalues of the Loschmidt matrix, see equations \eqref{rle} and \eqref{ele}. Shown is the lowest eigenvalue for a bulk system of size $N=200^2$ implemented using periodic boundary conditions, and the results for an open system using open boundary conditions (OBC) of size $N=60^2$. The shaded region is the region in which bulk eigenvalues exist. The upper panel shows the quench form the non-trivial to the trivial phase and no zero eigenvalues occur. The lower panel shows the quench from the trivial to the non-trivial phase and four eigenvalues become pinned to zero between successive critical times $t_c$, in agreement with the dynamical bulk-boundary correspondence.}
    \label{fig:bound}
\end{figure}

In figure \ref{fig:bound} the lowest eigenvalues of the Loschmidt matrix are shown for the two quenches considered and for systems with both open and periodic boundary conditions. As predicted by the dynamical bulk-boundary correspondence four zero eigenvalues occur between alternate critical times, but only for the quench into the non-trivial phase. The slow decay of the boundary return rate for the quench $\nu:0\to1$ is explained by the slow increase in the absolute value of the eigenvalues which were zero for $t_c<t<3t_c$. The discrepancy between the lowest eigenvalues for the open and periodic systems is caused by the existence of gapped one dimensional edge states which exist on the boundary of the open system, but which naturally do not occur for periodic boundary conditions.

\begin{figure}
    \centering
    \includegraphics[width=0.95\columnwidth]{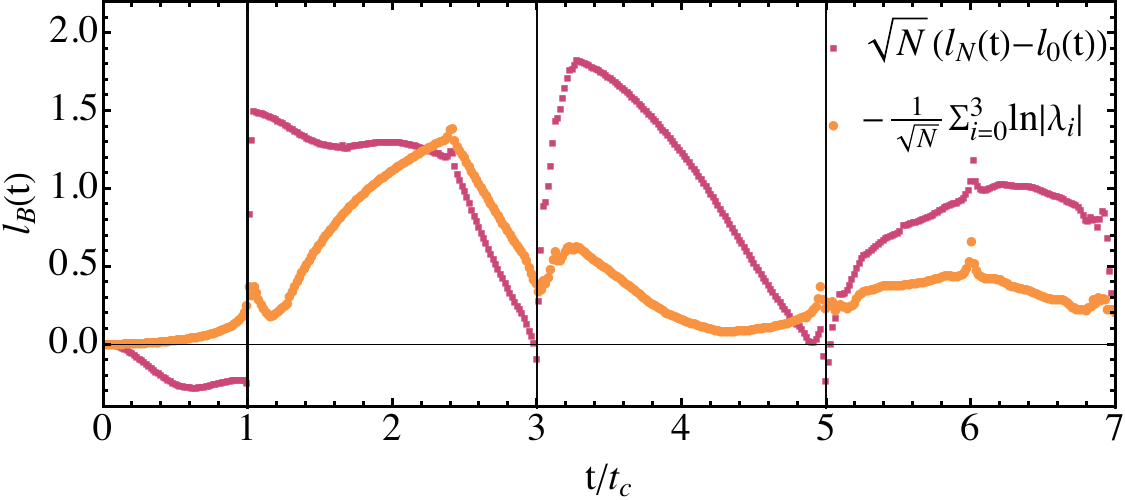}
    \caption{A comparison of the boundary contribution to the return rate extracted directly from $l_N(t)$ for $N=60^2$ and from the lowest Loschmidt eigenvalues, see equation \eqref{comp}.}
    \label{fig:plateau}
\end{figure}

We can compare the contribution to the return rate of just the lowest eigenvalues for the quench $\nu:0\to1$ to the boundary return rate. According to the dynamical bulk-boundary correspondence
\begin{equation}\label{comp}
    l_B(t)\approx \sqrt{N}\left(l_N(t)-l_0(t)\right)
    \approx-\frac{1}{\sqrt{N}}\sum_{i=0}^3\ln\left|\lambda_i(t)\right|.
\end{equation}
In figure \ref{fig:plateau} we compare these two quantities. Some qualitative agreement is visible, though at the system sizes we can achieve there is no quantitative agreement possible.

\section{Discussion and conclusions}

In this work we investigated dynamical quantum phase transitions and the dynamical bulk-boundary correspondence for an exemplary two dimensional higher order topological insulator. The relative simplicity of this model allows us to obtain higher quality data than has previously been obtained for DQPTs in more complicated higher order topological insulators. This simplicity also leads to a collapse of the area of Fisher zeroes onto lines of Fisher zeroes in the complex time plane, leading to DQPT behaviour which is characteristic of one dimensional rather than two dimensional topological models. Correspondingly we find that cusps occur in the return rate, and discontinuities occur in its derivative, at periodic critical times for quenches across a topological phase boundary.

We also see clear evidence of a dynamical bulk-boundary correspondence in the behaviour of the eigenvalues of the Loschmidt matrix. At critical times the spectrum of the Loschmidt matrix becomes gapless, and between alternate critical times there are ``in-gap'' eigenvalues pinned to zero, but only for quenches into the topologically non-trivial regime. These zeroes give rise to alternating plateaus in the boundary contribution to the return rate which we try to extract from a scaling analysis, comparing this to the boundary contribution at a specific system size and to the contribution form the zero eigenvalues. Here agreement is not perfect due to finite size errors. A systematic study of the dynamical bulk-boundary correspondence for different models and for quenches between a wider range of topological phases, and also of the origin of the Loschmidt zero eigenvalues, are interesting avenues to follow up.  

\acknowledgments
This work was supported by the National Science Centre (NCN, Poland) under the grant 2019/35/B/ST3/03625. Data can be found on Zenodo at 10.5281/zenodo.10571375.


%

\end{document}